\documentstyle[sprocl,epsfig]{article}

\def\Journal#1#2#3#4{{#1} {\bf #2}, #3 (#4)}

\newcommand{\be}{\begin{equation}}
\newcommand{\ee}{\end{equation}}
\newcommand{\bea}{\begin{eqnarray}}
\newcommand{\eea}{\end{eqnarray}}
\def\Pom{{\bf I\!P}}              %Pomeron symbol
\def\gsim{\mathrel{\rlap{\lower4pt\hbox{\hskip1pt$\sim$}}
    \raise1pt\hbox{$>$}}}         %greater than or approx. symbol
\def\lsim{\mathrel{\rlap{\lower4pt\hbox{\hskip1pt$\sim$}}
    \raise1pt\hbox{$<$}}}         %less than or approx. symbol

\def\NPB{{Nucl. Phys.} B}

\def\PLB{{Phys. Lett.}  B}

\def\ZPC{{Z. Phys.} C}

\begin{document}

\title{Diffractive DIS and QCD: past, present 
and future}

\author{M. Genovese}

\address{Istituto Elettrotecnico Nazionale Galileo Ferraris,
str. Cacce 91 , Torino, Italy} 
\maketitle

\section*{}

%\vskip 2cm
One of the most interesting discoveries at HERA has been
the observation of large 
rapidity--gap events, which give us the possibility of investigating
Diffractive DIS at small $x$, raising the hope  of relating
Regge theory with the calculations in perturbative QCD 
and, in particular, of understanding the nature of the 
dominant Regge trajectory at high energies, the "pomeron". 

A fundamental tool for these studies is the
colour--dipole scheme, introduced by Nikolaev and Zakharov 
\cite{NZ1} and by
Mueller\cite{Mueller}, which offers a unified approach of small $x$ 
inclusive DIS and Diffractive DIS.
This scheme derives from the observation that at small $x$ the
scattering process can be  conceived as the virtual photon fluctuating 
into a quark--antiquark pair, which interacts with the proton by the exchange 
of gluons. The $q \bar q$ fluctuation lives a time
$\Delta t \approx {1 \over x m_p}$ ($m_p$ is the proton mass)
much longer at small $x$ than the interaction time.
Therefore the transverse dimension of the pair can  be considered 
frozen during the scattering.
For what concerns the inclusive DIS, the interaction, at the lowest 
order, is given by the exchange of a gluon and the calculated cross 
section can be written as \cite{NZ1}:
\be
\sigma_{T,L}(\gamma^{*}N)=\int_{0}^{1} dz\int 
d\vec{\rho}
\,\vert\Psi^{\gamma^*}_{T,L}(z,\rho)\vert^{2}\sigma(\rho,x)
\label{eq:DIS}
\ee
in terms of a photon (with Transverse or Longitudinal polarization)
wave function
$\Psi^{\gamma^*}_{T,L}(z,\rho)$ (see \cite{NZ1} for its explicit form),
whose modulus squared gives the probability of 
producing a pair with transverse size $\rho$ and with a fraction $z$ of 
the photon light--cone momentum, and a cross section 
$\sigma(\rho,x)$, which can be directly related to the unintegrated 
glue distribution of the proton:
\be{
\sigma(x,\rho)= {4 \pi \alpha_{S} \over 3} \, \int
{ d\vec{k} \over \vec{k}^{4}}
(1- e^{i \vec{k} \cdot \vec{\rho} }) 
{d[x  g(x,k^{2})] \over d \ln \vec{k}^2}}
\ee
An important observation is that the dependences on the flavour of 
the pair and on the photon polarization are contained in $\Psi^{\gamma^*}$,
while $\sigma(\rho,x)$ is universal.
A similar analysis can be carried on for the DDIS, where one needs, to 
the lowest order, the exchange of two gluons.
The result can be written as:
\be
\left . {d\sigma_{T,L} \over dt} \right | _{t=0}= \int
{ dz \over 16 \pi} \int d\vec{\rho}
\,\vert\Psi^{\gamma^*}_{T,L}(z,\rho)\vert^{2}\sigma(\rho,x)^2
\label{eq:DDIS}
\ee
%where now  $\sigma(\rho,x)^2$ appears.
Incidentally, it must be noticed that in the literature Eq.s 
(\ref{eq:DIS},\ref{eq:DDIS}) are used
both in the momentum or in the dipole-size representations.
%according to which is the more appropriate for understanding the 
%particular case under consideration.

 Once $\sigma(\rho,x)$ is known, 
many processes can be calculated.
This allows to move from processes where one selects rather small
sizes (large $k^2$), and thus in the perturbative regime,
up to cases where rather large sizes are relevant and, therefore, there 
is a penetration into the non-perturbative region. The evaluation of
the contribution from the soft region requires some
modelling, as stating that the diagrams 
included in the calculation continue to be dominant and a 
parameterization of the unintegrated glue at small scales (see 
\cite{BGNPZ}). However, one can fix the relatively small number 
of parameters in some process and then use them for producing a large 
amount of predictions for other processes: the agreement with 
experimental data permits to check this ansatz.

The fact of selecting different scales varying the process leads
to the prediction of different effective pomeron intercepts:
this is 
a fundamental testable prediction of the colour--dipole scheme. 
For example, the vector meson  production amplitude can be written 
as\cite{VM}:
\be
< V | M | \gamma^* > = \int dz\int d\vec{\rho} \Psi_V^*(z,\rho) M(\rho )
\Psi^{\gamma ^*}_{T,L}(z,\rho)
\ee
where $M(\rho)$ is the amplitude for the scattering of a coloured dipole 
of size $\rho$ and $\Psi_V$ the vector meson wave function. 
The analysis of the integrand shows that one is selecting a scale
$ {q}^2 \approx (0.1 - 0.2) \cdot (Q^2 +M_V^2)$,
which for real photoproduction is rather small, leading to a small 
effective intercept, while increasing $Q^2$, it goes deeper and deeper
in the realm of perturbative QCD giving a rising effective intercept.
The variation of the effective intercept with the scale can be nicely observed 
looking at the experimental data on $\rho$, $\phi$ and $J/\Psi$ 
production\cite{VMexp}.

For what concerns the low-x inclusive DIS, the resummation of 
leading $log(1/x)$ in colour--dipole model \cite{NZ1,Mueller} 
 recovers the BFKL equation.
A plausible scenario for running BFKL equation is that we 
have a sequence of BFKL poles
plus a soft term 
\be
F_{2} (x,Q^2)\approx  \sum _i F_i(Q^2)  \cdot \left ( {1 \over x} \right ) 
^{\Delta_i}
+ F_{soft}(Q^2) \cdot \left ( {1 \over x} \right ) ^{\Delta_{soft}}
\ee
which, in a limited region of $x$, can be parameterized with
an effective intercept
$F_{2} \propto \left ( {1 \over x} \right ) ^{\Delta_{eff} }$.
However, at small $x$  one should also include 
unitarization effects, as the one coming from triple pomeron, which, 
in the colour--dipole scheme, derives from the
$q \bar q g $ Fock states of the photon \cite{a3p}.
Some first attempt of evaluating unitarization in colour--dipole model
have already appeared \cite{unit}, but a deeper 
analysis has yet to be performed. 
Anyway, experimentally the scale dependence of the effective intercept has been 
clearly observed \cite{DISexp}.

The most detailed analysis in the colour--dipole scheme has   been
carried out for DDIS, where various results have been 
derived in a good agreement with the  experimental data.

The study of the  transverse diffractive structure function 
at intermediate values of $\beta$,
where excitation of  the $q \bar q$ Fock component of the photon dominates, has been 
accomplished in \cite{lqqq} giving
\be
F_{T}^{D(3)}(x_{\Pom},\beta,Q^{2})\approx 
{\beta (1-\beta)^{2}(3+4\beta+8\beta^2)\over 6 m_{f}^{2}B_{d}(\beta)}
\cdot { e_{f}^{2} \over 12}
\cdot\left[\alpha_{S}({q}^{2})
G(x_{\Pom},{q}^{2} )\right]^{2}
\ee
For light quarks, a substantial penetration into the low 
scales is found, for the 
process is dominated by the scale ${q}^2 \approx 
{k^2 +m_f^2 \over (1 - \beta)}$.
However, for heavy quarks \cite{DDISL}, the large 
quark mass gives a perturbative scale, leading to a larger 
effective intercept than for the light quarks component (an interesting
breaking of Regge factorization, which we cannot discuss here, 
is found for charged currents DDIS, see \cite{CC}).

On the other hand, for a longitudinal polarization of the $\gamma^*$
the colour--dipole calculation \cite{DDISL}  shows that
one selects a hard scale $ {q}_L^2 \approx  {Q^2 \over 4 \beta}$,
independently on the flavour,
\be
F_{L}^{D(3)}(x_{\Pom},\beta,Q^{2}) \approx
{4\beta^{3}(1-2\beta)^{2}\over Q^{2}B_{3\Pom}}\cdot { e_{f}^{2}
\over 12} \left [\alpha_{S}({1\over 4}Q^{2})
G(x_{\Pom}, q_L^{2}) \right ]^{2}
\ee
One finds the same scale for the twist-4 component of the transverse 
structure function (which appears with a negative sign), 
studied in Ref. \cite{HT,BGNPZ}. 
The relevant point is that for these two last components
the ${1 \over Q^2}$ factor, due to the higher twist behaviour, is partially 
compensated by the growth of the glue with  
$ {q}^2_L$. $F^{D(3)}_2$ at large $\beta$
comes  from the sum of all these terms, and at $\beta \gsim 
0.8$ is dominated by $F^{D(3)}_L$ even at relatively large $Q^2$ (see 
figure 1). 
Thus at large $\beta$ one cannot apply to diffractive structure 
function the usual QCD evolution.
The different $x_{\Pom} $ dependences of these different components
and their comparison with experimental data \cite{H1,ZEUS} are
presented in figure 1.

\begin{figure}[t] \label{FIGUR1}
\begin{center}
\mbox{\epsfig{file=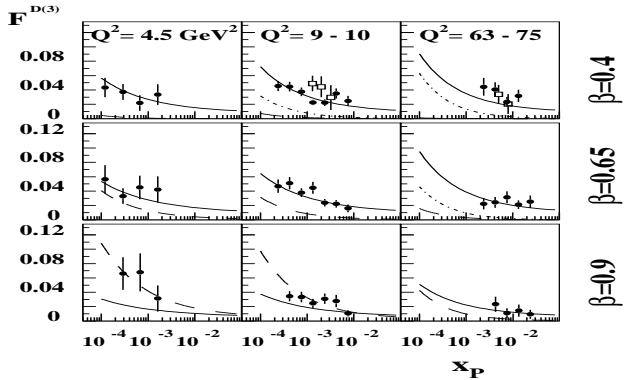,width=0.8\textwidth,height=5.5cm}}
\end{center}
\caption{Colour-dipole calculation (see \protect\cite{CC}) of $x_{\Pom}$ dependences 
for $F^{D(3)}_T$ (solid), $F^{D(3)}_L$
(dashed) and $5 \cdot F^{D(3)}_T(c) $ (dot--dashed). H1\protect\cite{H1}
(circles)  and ZEUS\protect\cite{ZEUS} (stars) $F_2^{D(3)}$
data are shown for comparison.} 
\end{figure}
For what concerns the small $\beta$ region, it is dominated 
by excitation of the $q \bar q g$ Fock states of the photon, calculated 
in Ref. \cite{NZ1,DDISTP}:
in this case an approximate factorization and the QCD evolution 
are recovered. There is still a penetration in 
the small momenta region, albeit less deep than for the light--quark 
transverse $q \bar q$ component, thus the effective intercept is 
larger. Altogether  the prediction of this scheme is that 
a fit of the form $F^{D(3)} \propto 
x_{\Pom}^{-\delta}$ is expected to give a large exponent at large 
$\beta$ ($\delta \approx 0.3$), where $F_L^D$
dominates, which decreases
in the region dominated by  the transverse $ud$
$q \bar q$ component reaching a minimum for $\beta \approx 0.6$,
where $\delta \approx 0.15$, and increases again in the low $\beta$ 
region.
The available DDIS experimental data do not yet have  a sufficient
statistics for fitting these exponents with different $\beta$ bins
(and excluding the regions were large contributions from other Regge
trajectories are expected. For a first attempt see \cite{Fiore}).
However, altogether, 
predicting the variation of the effective pomeron intercept
with the scale is one of the most relevant results obtained in the
colour--dipole scheme. In figure 2 the measured effective intercepts 
\cite{VMexp,DISexp,H1,ZEUS} in different processes
are shown in function of the effective scales (see \cite{NNNDIS} for
definitions). The scale dependence is clearly observed, although the 
available fits did not explore the predicted $\beta$-dependence of the 
effective intercept.
\begin{figure}[t] \label{FIGUR2}
\mbox{\epsfig{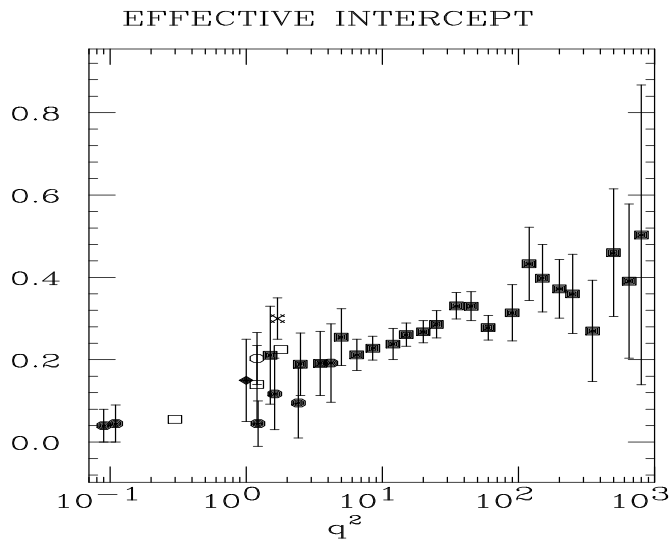}}
\caption{Experimental effective pomeron intercepts in function of the 
scale $q^2$, for rho (filled circle), $\phi$ (empty quadrangle) and 
$J/\Psi$ (cross) production \protect\cite{VMexp}, 
ZEUS \protect\cite{ZEUS} (rhombus) and H1 
\protect\cite{H1} (empty circle) DDIS and H1 inclusive DIS \protect\cite{DISexp} 
(filled quadrangle)} 
\end{figure}

In conclusion, the colour--dipole model has been applied to a large
number of different processes, giving a unified treatment of low $x$ 
inclusive and diffractive DIS \footnote{The model has also been applied 
successfully to the study of nuclear structure functions, see Ref. 
\cite{shad}}. For different processes one selects different 
scales,  with the possibility  of studying the
transition  between perturbative  and non-perturbative QCD.
 Up to now a good agreement with available experimental data  has been
obtained;  more refined  tests of this scheme will soon be possible. 
\break

\end{document}